\documentclass[aps,prl,showpacs,float]{revtex4-1}
\usepackage{graphicx}
\begin{document}
\title{Comment on ``Computational method for the quantum Hamilton-Jacobi equation: Bound states in one dimension" [J. Chem. Phys. 125, 174103 (2006)]}
\author{Mario Fusco Girard}
\affiliation{Department of Physics ''E.R. Caianiello'',\\University of Salerno \\and \\Gruppo Collegato INFN di Salerno,\\Via Giovanni Paolo II, 84084 Fisciano (SA), Italy}

\begin{abstract}
 Some difficulties, both numerical and conceptual, of the method to compute one dimensional wave functions by numerically integrating the quantum Hamilton-Jacobi equation, presented  in the paper mentioned  in the title,  are analyzed. The origin of these difficulties is discussed, and it is shown how they can be avoided by means of another approach based on different solutions of the same equation. Results for the same potentials, obtained by this latter method are presented and a comparison is made.
\end{abstract}

\maketitle

\section{}
Few years ago,  Chou and Wyatt presented "an accurate computational  method for the one-dimensional quantum Hamilton-Jacobi" [1]. By means of this approach, ``bound state wave functions are synthesized"… and ``accurately obtained".  In the paper, the method was applied to two solvable examples, the harmonic oscillator and the Morse potential, with ``excellent agreement with the exact analytical results", so that the proposed  procedure ``may  be useful for solving similar quantum mechanical problems".
However, the presented method faces a series of difficulties, both  numerical and conceptual, as discussed in the following.
The method exploits the approach proposed by Leacock and Padgett [2,3], who renewed the old interest for the Quantum Hamilton-Jacobi Equation (QHJE) [4,5].  The starting point, as in the well known Wentzel-Kramers-Brillouin method, is  the search for a solution of the form
\begin{equation}
\psi(x,E)=e^{{i\over \hbar} W_L(x,E)}
\end{equation}	
of the time-independent Schroedinger Equation (SE) for a particle of mass m in a potential V(x).
\begin{equation}
-{\hbar ^2\over 2m} {d^2\psi\over dx^2}=\left[E-V(x)\right]\psi\ .
\end{equation}    

By substituting Eq. (1)  in (2),  the time independent quantum Hamilton-Jacobi equation results:
\begin{equation}
{1\over 2m}\left({dW_L\over dx}\right)^2 - {i\hbar \over 2m}{d^2W_L\over dx^2}=E - V(x)\ .
\end{equation}  

In Eqs. (1) and (3)  $W_L(x,E)$ is a quantum characteristic function or (reduced) action of the particle ( the reason for the pedices will be clear soon).  Its derivative
\begin{equation}
p_L (x,E)={\partial W_L (x,E)\over \partial x}	
\end{equation} 
is called  the Quantum Momentum Function (QMF), in terms of which the QHJE is rewritten as
\begin{equation}
{1\over 2m} p_L (x,E)^2+{\hbar\over 2mi}{\partial p_L (x,E)\over \partial x} = E- V(x)\ .	
\end{equation} 

To complete the definition of  $p_L(x,E)$,  a ``physical boundary condition" is imposed
\begin{equation}
\lim_{\hbar\to 0}p_L(x,E) = p_C (x,E)
\end{equation}		 
where $p_C(x,E)$ is  the particle's classical momentum
\begin{equation}
p_C (x,E) = \pm {\sqrt{2m(E-V(x))}}\ .
\end{equation}                   
From Eq. (1) ,

\begin{equation}
W_L(x,E) = {\hbar \over i} \ln \psi(x,E)\ .
\end{equation} 		

Therefore
\begin{equation}
p_L(x,E) = {\hbar \over i\psi(x,E)}{\partial \psi (x,E)\over \partial x}\ .
\end{equation}			
From the last equation, it is clear that each  node of the wave function corresponds to a first order pole in $p_L(x,E)$. In [2,3], it is shown how the study of the poles of the QMF in the complex x-plane allows to obtain the exact quantization condition for the one dimensional case, without solving the SE or the equivalent QHJE.

The procedure to numerically integrate equation (5) proposed in [1] starts from the observation that the poles in the QMF prevents the use of standard methods. Chou and Wyatt therefore adopt a particular technique, proposed by Schiff and Shnider [6], i.e. a third order Moebious integrator method. This obviously complicates the integration, with respect to usual numerical procedures.

From the knowledge of the QMF, the wave function can be obtained by numerically integrating Eq. (9). For the ground state, the integration offers no problem, but for the excited states, the poles in the integrand prevents the use of ordinary integration procedures, and another special method  is adopted, i.e. the ``antithetic cancellation technique". This method requires to use  ``mirror sets" of points, symmetrically disposed with respect to the nodes, in order to cancel the divergences when summing the integrands. This in turn requires  the ``a priori"  knowledge of the position of the nodes, or at least an estimate of their positions, to be improved by means of a particular numerical procedure, as done in Ref. [1].

Obviously, also this second problem complicates the integration, and the complexity increases with the quantum number $n$.

However, the most  important critical points of the method proposed in [1] are of conceptual type. In order to start the numerical procedure to compute the QMF, it is necessary to choose an initial value for $p_L(x,E)$. In Ref. [1] the authors use the value of the classical momentum in a point  chosen in the classically forbidden region. One problem is that according to Eq. (6), the identification of the QMF with the classical momentum should only hold in the limit for $\hbar \to 0$, and is not justified when computing an exact quantum wave function. Actually, it is  immediate to verify that in the cases investigated in Ref.  [1],  $p_L$  is very different from the classical momentum $p_C$, also in the classically forbidden regions, where both are imaginary quantities. The problem is worse in the classically allowed region, where the classical momentum is a real function of x, while the QMF is a purely  imaginary quantity, and continues to be imaginary also in the limit for $\hbar\to 0$, as clearly seen from eq. (9). This difficulty is noted in Ref. [1], but no solution  is proposed. Finally, in the classical limit $\hbar\to 0$, the poles of $p_L$ coalesce, originating a segment of the real $x$-axis containing a denumerable infinity of polar singularities: in this  limit instead $p_L$, according to the condition (6), should go into the perfectly regular classical momentum.

The difficulties above originate from the choice to represent the {\it effective, real}  wave function in the form given in Eq. (1). This form is convenient for the analytical study of the QHJE in the complex plane, as in [2,3], but is not appropriate to numerically integrate the same equation on the real $x$-axis or to perform the classical limit for $\hbar\to 0$.  Indeed, it is clear from  Eq. (8) that if $\psi$ is the effective wave function, its nodes impose the presence of branch point singularities to the quantum characteristic function $W_L$, which in turn give poles to the QMF; moreover the QMF for a real wave function is a purely imaginary quantity along all the x-axis. The singularities for $W_L$ and  $p_L$  and  the imaginary character of this latter quantity are preserved in the classical limit;  instead in that limit, the QHJE transforms into the classical Hamilton-Jacobi, whose solution W0 and the corresponding classical momentum $p_C$ are regular real functions of $x$, in all the classically allowed region.

An approach which avoids all these difficulties was presented in [7].  In the classically allowed region, through Eq. (1)
 one constructs an {\it auxiliary, complex} solution $\psi_M$ of the SE at the energy E, by means of a solution
\begin{equation}
W_M  = X(x)+i Y(x)\ ,
\end{equation}       
of the QHJE,  different from $W_L$, and continuous in all the classical region. This is possible because, as shown in [7], to the same wave function it corresponds a whole one-parameter family of solutions of the QHJE. Then, as in the ordinary WKB method,  the effective, real wave function $\psi$ is obtained by combining the auxiliary solution $\psi_M$ with its complex conjugate $\psi_M^*$. In [7] it is shown that the effective wave function can be written in the WKB-like form in the classical region:
\begin{equation}
\psi(x,E) ={A\over \sqrt{|X'(x)|}}\sin\left[{X(x)\over \hbar}+{\pi\over 4}\right]\ ,
\end{equation}        
Here $X(x)$ is the real part of the quantum action $W_M$, $X'(x)$ is its derivative, and A is a constant.				
In the classically forbidden regions, the wave function is instead represented by
\begin{equation}
\psi(x,E)=B_i e^{-Y_i(x,E)/\hbar}\ ,
\end{equation}     
where $Y_i (x, E)$ are suitable solutions of the QHJE in the forbidden regions ( the quantum characteristic function is purely imaginary there), are $B_i$ are constants.  Differently from the approximate WKB representation, the representation given by Eqs. (11) and (12) is exact along all the real $x$-axis, and exactly reproduces the wave function also  near the turning points, where the WKB method notoriously fails.
The nodes of the effective wave function are simply due to the interference of the two terms, $\psi_M$ and $\psi_M^*$. In the classical region the quantities $W_M$ and the corresponding QMF, $p_M$, are smooth complex functions without branch points or poles, respectively,  so all the numerical difficulties of the procedure presented in Ref.[1] are absent and standard numerical integration methods are applicable. There is no need of the preventive knowledge of the poles' positions. Moreover, in the classical limit both $W_M$  and $p_M$  go into the  corresponding classical quantities,  due to  their imaginary parts being proportional to $\hbar$. Finally, the real quantum quantity to be compared with the classical momentum is not $p_L$ but the derivative $X'$ of the real part $X$ of the quantum action $W_M$.
In [7] the method is described in detail and applied to the harmonic oscillator and to the radial motion in a Coulomb potential. The method is general and for all one-dimensional potential so far investigated gives the wave functions with great accuracy. The details of the procedure are presented in [7] and will not repeated here, where for comparison we present instead the results for two cases already studied in Ref. [1], i.e. the construction of a wave function for the harmonic oscillator and another for the Morse potential.
Figures 1-3  refer to the harmonic oscillator, while Fig. 4  is for the Morse potential. Units and parameters' values are the same as in Ref. [1]. In Fig. 1 the real part X(x) of the quantum characteristic function $W_M(x)$ for the state $n=2$ of the harmonic oscillator in the classically allowed region is reported  together with the corresponding classical quantity $W0(x)$.  Fig. 2 shows the derivative $X'(x)$ and the classical momentum $p_C(x)=W0'(x)$. In Fig. 3 are plotted the quantities ${1\over \sqrt{|X'(x)|}}$, $\sin\left[{X(x)\over \hbar}+{\pi\over 4}\right]$ and finally their product, which gives according to Eq. (11) the  eigenfunction $\psi$ in the classical region.  For simplicity the left and right exponential tails in the forbidden regions are not reported. This wave function accurately reproduces the well known $n=2$ state of the oscillator along all the real $x$-axis.
 Fig. 4 refers to the $n=2$ state of the Morse oscillator. In the upper box the solutions of the QHJE are plotted; the continuous line gives the real part $X(x)$ of the quantum action $W_M$ in the classical region, between the two turning point. The dashed curves report the imaginary parts $Y(x)$ of the action in the forbidden regions. In the lower box,  the resulting  wave function is presented which also in this case accurately reproduces the corresponding  solution of the SE along all the $x$-axes. All the results in the figures are obtained by standard integration methods.

\begin{figure}[h]
 \includegraphics[scale=0.8]{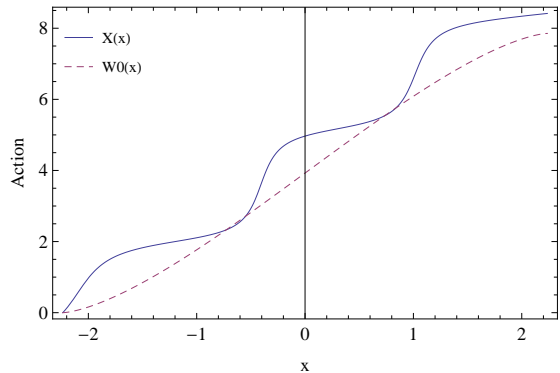}%
 \caption{\label{}The real part $X(x)$ of the quantum characteristic function $W_M(x)$ (continuous line) and the classical action $W0(x)$ (dashed line) for the $n=2$ state of the harmonic oscillator.}
 \end{figure}

\begin{figure}[h]
 \includegraphics[scale=0.8]{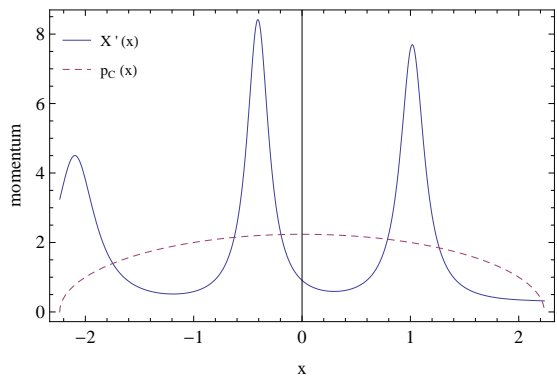}%
 \caption{\label{}The derivative $X'(x)$ (continuous line) and the and the corresponding classical momentum $p_C(x)$ (dashed line) for the $n=2$ state of the harmonic oscillator.}
 \end{figure}

\begin{figure}[h]
 \includegraphics[scale=0.8]{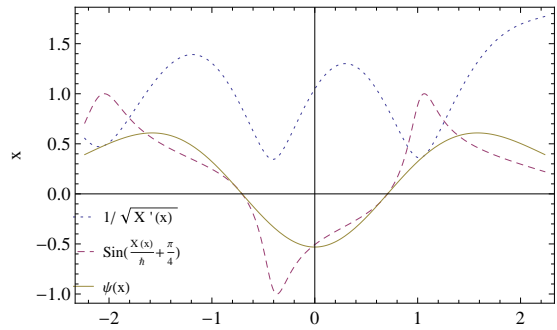}%
 \caption{\label{}The quantities  $1\sqrt{|X'(x)|}$ (dotted line), $\sin [{X(x)\over \hbar}+Pi/4]$ (dashed line) and their product (continuous line), which according to Eq. (11) represents the wave function in the classical region, for the $n=2$ state of the harmonic oscillator.}
 \end{figure}

\begin{figure}[h]
 \includegraphics[scale=0.8]{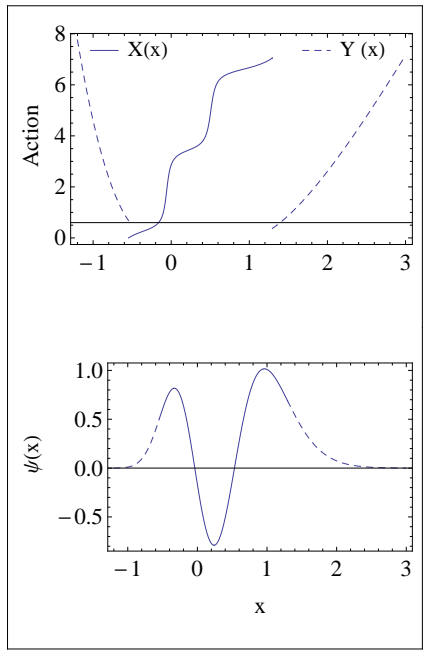}%
 \caption{\label{}Upper box: the solutions $Y_1(x)$ and $Y_3(x)$ of the QHJE in the two classically forbidden regions (dashed lines), and  the real part of the quantum action, $X(x)$ in the classically allowed region (continuous line) for the $n=2$ state of the Morse potential. Lower box: the corresponding eigenfunction, constructed in the forbidden regions from the solutions of the QHJE plotted in the upper box by means of Eq. (12) (dashed lines) and by means of Eq. (11) for the classical region (continuous line).}
 \end{figure}

\end{document}